# TEVATRON ACCELERATOR PHYSICS AND OPERATION HIGHLIGHTS

A. Valishev for the Tevatron group, FNAL, Batavia, IL 60510, U.S.A.

*Abstract*

The performance of the Tevatron collider demonstrated continuous growth over the course of Run II, with the peak luminosity reaching $4\times10^{32}$ cm$^{-2}$ s$^{-1}$, and the weekly integration rate exceeding 70 pb$^{-1}$. This report presents a review of the most important advances that contributed to this performance improvement, including beam dynamics modeling, precision optics measurements and stability control, implementation of collimation during low-beta squeeze. Algorithms employed for optimization of the luminosity integration are presented and the lessons learned from high-luminosity operation are discussed. Studies of novel accelerator physics concepts at the Tevatron are described, such as the collimation techniques using crystal collimator and hollow electron beam, and compensation of beam-beam effects.

## COLLIDER RUN II PERFORMANCE

Tevatron collider Run II with proton-antiproton collisions at the center of mass energy of 1.96 TeV started in March 2001. Since then, 10.5 fb$^{-1}$ of integrated luminosity has been delivered to CDF and D0 experiments (Fig. 1). All major technical upgrades of the accelerator complex were completed by 2007 [1]. Nevertheless, the collider performance continues to exhibit significant growth: in 2010 2.47 fb$^{-1}$ of luminosity was integrated, and the peak luminosity reached $4\times10^{32}$ cm$^{-2}$s$^{-1}$ (integrated luminosity by year is listed in Table 1). In FY2011, 1.2 fb$^{-1}$ was accumulated until the time of this report (March 2011), well on track to delivering the planned 2.6 fb$^{-1}$ by the end of the year, which would make the Run II total close to 12 fb$^{-1}$. Table 2 lists the main collider parameters achieved during Run II as compared to the design goals.

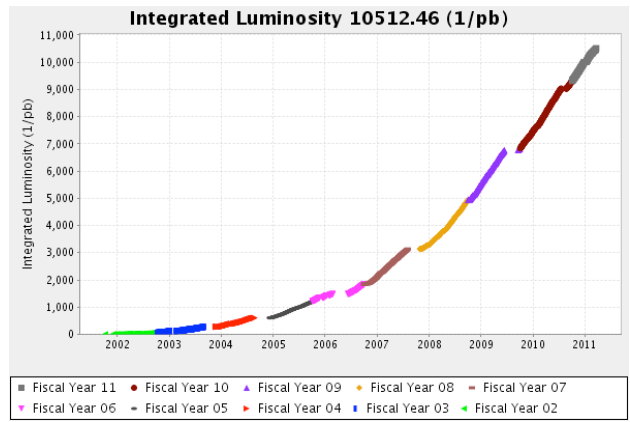

Figure 1: Run II integrated luminosity by fiscal year.

Table 1: Integrated luminosity performance by fiscal year.

|  | FY07 | FY08 | FY09 | FY10 |
|---|---|---|---|---|
| Total integral (fb$^{-1}$) | 1.3 | 1.8 | 1.9 | 2.47 |

Until the middle of calendar year 2009, the luminosity growth was dominated by improvements of the antiproton production rate [2], which remains stable since. Performance improvements over the last two years became possible because of implementation of a few operational changes, described in the following section.

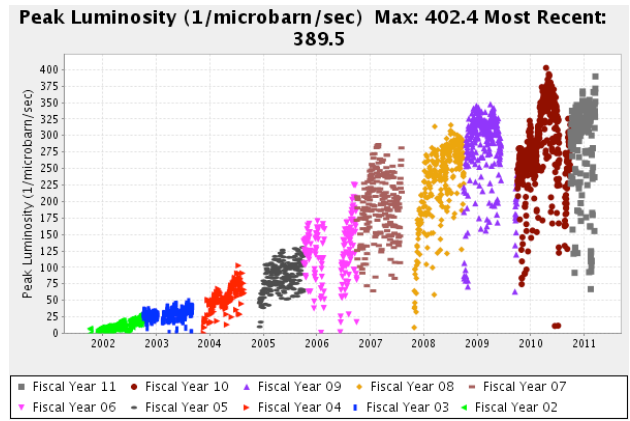

Figure 2: Run II peak luminosity by fiscal year. $\mu$b$^{-1}$/s = $10^{30}$ cm$^{-2}$s$^{-1}$.

Table 2: Main collider parameters

|  | Design | Achieved |
|---|---|---|
| Antiproton production rate ($10^{10}$/h) | 32 | 22 |
| Stack to HEP $\bar{p}$ transfer efficiency | 80% | 83% |
| Initial luminosity ($10^{32}$ cm$^{-2}$s$^{-1}$) | 2.9 | 4.0 |
| HEP store duration (h) | 15 | 15 |
| Shot setup time (h) | 2 | 1 |
| Store hours per week (h) | 97 | 120 |
| Luminosity integral per week (pb$^{-1}$) | 55 | 73 |

## OPERATIONAL IMPROVEMENTS

### Antiproton Storage

With the achievement of stable high stacking rate of antiprotons in the Accumulator [3], and improvement of the beam transfer to the Recycler [4], beam lifetime during storage in the Recycler became an essential factor. It was determined that the beam brightness is limited by a transverse instability [5]. Streamlining of the RF manipulation procedures and improved vacuum allowed to achieve the lifetime of 200-400 h up to the intensity of $5.25\times10^{12}$. This resulted in 3-5% improvement of the Recycler storage efficiency, which is now typically 93-95%.

*Proton Scraping in Main Injector*

High brightness of the proton beam delivered to the collider is essential for achievement of high luminosity. Moreover, large amount of halo particles elevates the losses along the collider cycle and increases the possibility of a quench. Hence, a momentum scraping of the proton beam was implemented in the Main Injector at the injection energy (8 GeV). This is realized as an orbit bump close to collimator at a high dispersion location. The effect of this procedure was a 3-4% increase of the initial luminosity (Fig. 3).

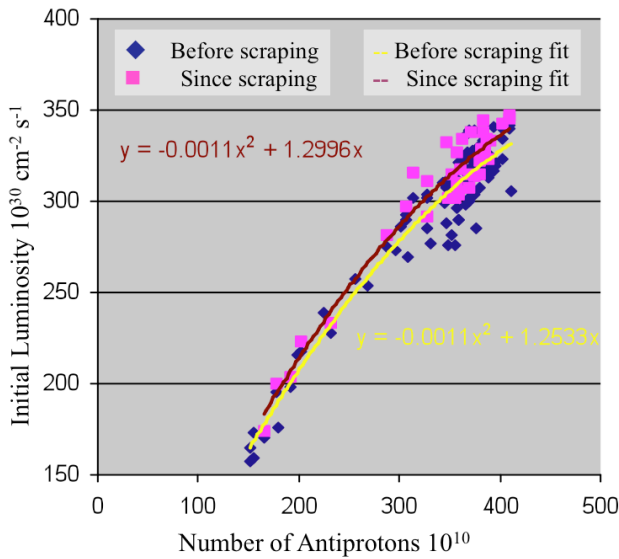

Figure 3: Initial luminosity vs. the number of antiprotons. Comparison of operation before and after implementation of proton scraping in the Main Injector.

*Tevatron Performance and Stability*

In the Tevatron collider (see Table 3 for main parameters), the beams of protons and antiprotons move along helical separated orbits in a common vacuum chamber. Because of this, each bunch experiences 70 parasitic (long-range) collisions around the ring in addition to the two main head-on collisions producing luminosity. After optimization of the helical orbits in 2007, the particle losses during injection and acceleration stay at an acceptable level: about 5% of protons are lost during the pbar injection, and 2% of both species is lost on acceleration. Because the intensity loss on the ramp occurs at lower energy, this typically does not lead to a magnet quench.

The low-beta squeeze is a much more critical stage of the collider cycle. Two significant changes occur simultaneously during the 120-second step: the value of $\beta^*$ is gradually decreased from 1.5 m to 0.28 m, and the helical orbits change their shape and direction of rotation from the injection to the collision configuration. The latter poses a serious limitation since the beams' separation at several long-range collision points briefly (during approx. 2 s) decreases from $6\sigma$ to $\sim 2\sigma$. At this moment a sharp spike in losses localized at the low-beta regions is observed. These losses often damage sensitive detector equipment, but more importantly cause superconducting magnet quenches and, consequently, loss of store. Figure 4 presents the categorization of Tevatron magnet quenches during the period between October 2007 and March 2011. As can be seen, out of the total number of 154 there were 32 beam loss induced quenches during the low-beta squeeze compared to only 5 during acceleration, 3 during halo removal, and 4 in high-energy physics (HEP) runs. Quenches during HEP stores occur at random times, quite often towards the end of the store. Because the cryogenic recovery time can be as short as 3 hours, this does not impact the luminosity integration. On the contrary, a quench during the low-beta squeeze means the loss of the entire antiproton stash, which requires lengthy replenishment, and results in about 8 pb$^{-1}$ effect on the luminosity integral. Thus, the 32 quenches in the squeeze account for a month of collider operation over the period of 3 years, or 3%. At high luminosity the probability of a quench increases and the statistics for 2010 shows the loss of 4% of luminosity owing to quenches in the squeeze.

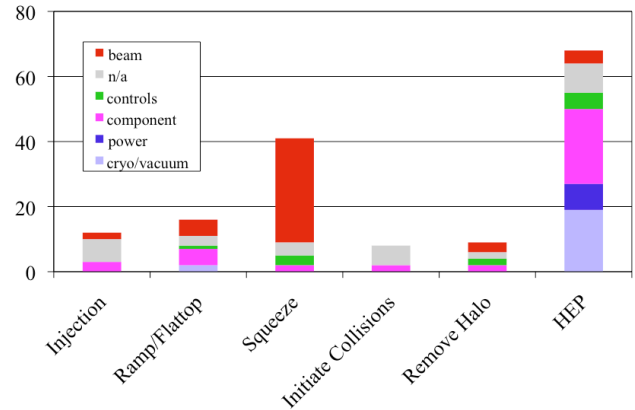

Figure 4: Categorization of Tevatron magnet quenches. Data between October 2007 and March 2011.

Limited success in reduction of the loss spike was achieved in 2009 by fine-tuning the helical orbits and lowering the betatron tune chromaticity [6]. Nevertheless, the issue required constant attention and a lot of time and effort was spent keeping the optimal configuration. A decisive solution was introduced by implementing collimation at the top energy and maintaining stable orbit around the collimator during the squeeze. A single proton collimator is efficient in absorbing the halo particles and shielding the interaction regions. Since the collimation at flat-top was implemented in store 8330 on December 2, 2010, the losses at experiments were reduced by 2 orders of magnitude and no beam-induced quenches occurred in 114 HEP stores. Before this modification, one in every 30 stores would result in a quench.

Table 3: Machine and beam parameters

| | |
|---|---|
| Number of bunches | 36 |
| Protons per bunch ($10^{11}$) | 2.9 |
| Antiprotons per bunch ($10^{11}$) | 1.0 |
| Proton emittance (95% normalized, $\mu$m) | 18 |
| Antiproton emittance (95% normalized, $\mu$m) | 8 |
| Proton bunch length (m) | 0.55 |
| Antiproton bunch length (m) | 0.45 |
| Number of IPs | 2 |
| Beta-function at IP (m) | 0.28 |
| Betatron tunes ($Q_x, Q_y$) | 20.583, 20.585 |
| Beam-beam parameter | 0.024 |

## *Operations Strategy*

The development of individual elements of the collider complex would not materialize in the luminosity integral without the careful planning and operation modeling by the Run Coordinators' team. A model of the entire complex of accelerators was built and used for optimization of the day-to-day collider performance, and to determine the possible areas of improvement [7].

The model includes the following key components:
- Antiproton transmission efficiency and time between the Accumulator and the Tevatron HEP store as a function of various parameters.
- Stacking rate as a function of the stack size in the Accumulator.
- Antiproton lifetime in the Recycler as a function of the stash size.
- Model of the Tevatron initial luminosity (including saturation due to beam-beam effects) and luminosity decay.
- Tevatron shot setup time.

The strategy of the luminosity optimization assumes stable machine parameters and continuous repetition of a pre-set cycle consisting of a series of antiproton transfers between the Accumulator and the Recycler, simultaneously with the HEP operation; and then a shot setup. By adjusting the frequency of antiproton transfers between the Accumulator and the Recycler one can maximize the number of pbars available for the next shot to the collider. The length of HEP store sets the stash size in the Recycler and, consequently, the initial luminosity. In Figure 5 the calculated weekly luminosity integral and stash size are plotted as a function of the (repeated) store duration for the assumed pbar accumulation rate of $21\times10^{10}$/hour. In Fig. 6 the actual luminosity integration during a typical week is shown.

The luminosity model is also instrumental in working out scenarios of various failures allowing to minimize the lost time, in setting proper time of machine development studies and accelerator maintenance accesses.

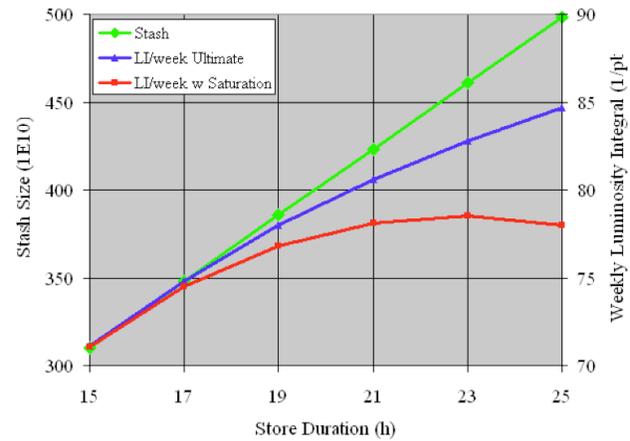

Figure 5: Luminosity integral per week (red, blue) and initial number of antiprotons (green) as a function of store length. Blue line assumes no beam-beam effects in the Tevatron, red uses an empirical model.

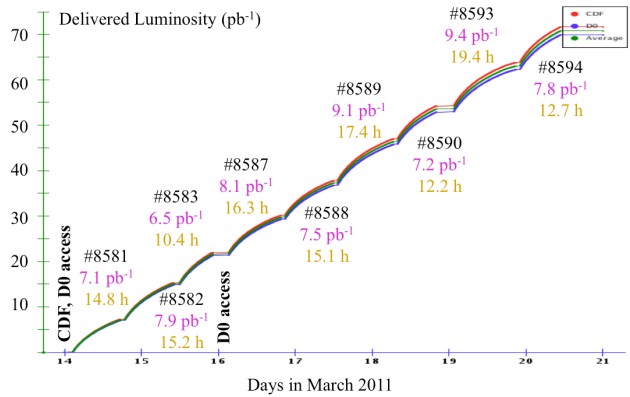

Figure 6: Luminosity integration during a typical 'good' week.

## ACCELERATOR PHYSICS STUDIES

Stable operation of the collider complex allows to develop and perform studies of novel accelerator physics techniques and technology that can benefit future machines. A plan exists to perform a wide range of beam physics experiments at the end of Run II [8]. The range of topics is wide and includes studies of particle diffusion due to beam-beam effects, intra-beam scattering, electron cloud, coherent beam-beam modes, luminosity leveling and beam instrumentation.

Some of the studies have been on going, taking advantage of parasitic (concurrent with HEP) operation or short dedicated periods of time, usually at the end of physics stores to minimize the impact on luminosity integration. These include the studies of head-on beam-beam compensation with the Tevatron electron lenses (TELs), and collimation of high-energy protons and antiprotons with crystal collimator and hollow electron beams.

### Head-on Beam-Beam Compensation

Mitigation of the head-on beam-beam effect in antiproton beams by compensation of the proton space charge using low energy electron beam lens has been proposed to improve the antiproton lifetime in collisions [9]. Two electron lenses (TEL-1 and TEL-2) have been built and installed at the Tevatron. TEL-1 is used in collider operation for cleaning of the abort gap particles, while the second lens can be used for studies. During 2009, TEL-2 was equipped with a Gaussian profile electron gun and in 2010 two attempts to demonstrate head-on beam-beam compensation in protons were performed without success [10]. However, these experiments provided valuable data that will be useful for development of simulation codes and for the beam-beam compensation project at RHIC [11].

### Collimation with Hollow Electron Beams

Low-energy hollow profile electron beam can be used for cleaning of halo particles in high-energy hadron machines. Compared to the conventional collimation systems, this technology is advantageous because the electron beam cannot be damaged by the high power circulating beam, and can be placed very close to the beam core. A hollow electron gun was developed, manufactured and installed in TEL-2 during the summer shutdown in 2010. Experiments with the hollow e-beam collimator started in 2011, and significant progress was achieved in both understanding the physics of this type of collimation, and demonstration of collimation [12].

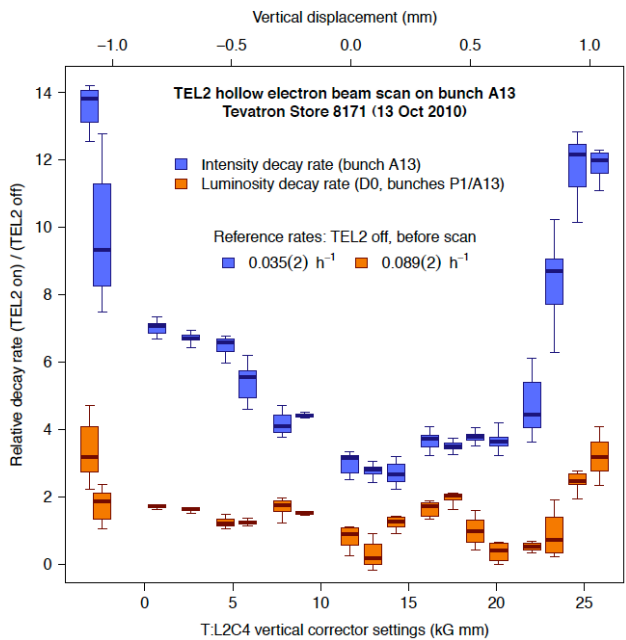

Figure 7: Ratio of luminosity and beam intensity decay rate during a position scan with the hollow e-beam collimator.

Figure 7 illustrates the effect of collimation on a single antiproton bunch. The luminosity decay of the affected bunch is the same as of the control bunch, while the intensity decay is faster, which means the e-beam collimator is removing particles with large amplitudes without touching the beam core.

### Collimation with Bent Crystals

Deflection of halo particles by bent crystals may improve the collimation efficiency. The T980 experiment at the Tevatron successfully demonstrated channeling of particles during many HEP stores [13,14]. Several types of crystals were studied, including the advanced multi-strip design.

## ACKNOWLEDGMENTS

The results reported in this paper were achieved because of hard work and dedication of many people at Fermilab. The author would like to thank J. Annala, C. Gattuso, V. Lebedev, R.S. Moore, L. Prost, V. Shiltsev, G. Stancari, D. Still, for valuable input and help in preparation of this work.